\documentclass[11pt]{article}

\usepackage[preprint]{acl}

\usepackage{times}
\usepackage{latexsym}
\usepackage[T1]{fontenc}
\usepackage[utf8]{inputenc}
\usepackage{inconsolata}
\usepackage{url}
\usepackage{booktabs}
\usepackage{amsfonts,amsmath,amssymb}
\usepackage{nicefrac}
\usepackage{microtype}
\usepackage{graphicx}
\usepackage{multirow}
\usepackage{fvextra}
\usepackage{xcolor}
\usepackage{colortbl}
\let\origtexttt\texttt
\renewcommand{\texttt}[1]{{\upshape\origtexttt{#1}}}
\usepackage{tikz}
\usepackage{pgfplots}
\pgfplotsset{compat=1.18}
\usetikzlibrary{positioning,arrows.meta,fit,backgrounds,calc,shapes.geometric}

\definecolor{cModel}{HTML}{2F5D8A}
\definecolor{cModelFill}{HTML}{E8F0F8}
\definecolor{cCode}{HTML}{6B6B6B}
\definecolor{cCodeFill}{HTML}{F2F2F2}
\definecolor{cAccent}{HTML}{B5532C}
\definecolor{cTrackFill}{HTML}{ECECEC}
\definecolor{cDocA}{HTML}{2F5D8A}
\definecolor{cDocB}{HTML}{C9803E}
\definecolor{cOurs}{HTML}{2F5D8A}
\definecolor{cGPT}{HTML}{C9803E}
\definecolor{cZephyr}{HTML}{8FB3D1}
\definecolor{cFollow}{HTML}{BDBDBD}
\definecolor{cAblA}{HTML}{7FA7CB}
\definecolor{cAblB}{HTML}{C9803E}
\definecolor{cAblC}{HTML}{BDBDBD}

\newcommand{\sys}{DMReranker}
\hypersetup{colorlinks=true, linkcolor=cModel, citecolor=cModel, urlcolor=cModel}

\title{Every Clause Counts: Instruction-Following Reranking\\ by Multiplying Small, Answerable Questions}

\author{S. M. Mahbub Uddin Ahmed \\
  United International University \\
  Dhaka, Bangladesh \\
  \texttt{sahmed221068@bscse.uiu.ac.bd}
  \And
  Abdullah Al Musaddiq \\
  Islamic University of Technology \\
  Gazipur, Bangladesh \\
  \texttt{musaddiq@iut-dhaka.edu}}

\begin{document}
\maketitle

\begin{abstract}
People often attach instructions to their search queries, asking for pages written for beginners, documents in a certain language or official sources only. Most rerankers were trained to judge topical relevance and treat such instructions as a few extra words. On the InfoSearch benchmark, even GPT-4o used as a list-wise reranker reaches an average WISE (Weighted Instruction Sensitivity Evaluation) of only 33.5. We present \sys{} (Decision-Model Reranker), a zero-shot reranker that splits an instructed query into a topic and a short list of typed clauses. A small judge model answers one narrow question per clause, and the answer probabilities are multiplied into a single score. Nothing is trained, and the configuration was frozen before the test runs. Reranking the same top-100 candidates as the published baselines, \sys{} reaches a macro WISE of 50.9 over all six InfoSearch dimensions while keeping original-mode nDCG@10 at 82.9. It trails GPT-4o on the stricter SICR (Strict Instruction Compliance Ratio) metric, 19.2 against 32.1. We show that this gap follows from how a product of probabilities behaves when SICR compares raw scores across runs. Computed from ranks, our SICR is 67.0. On the same judge, splitting the instruction raises WISE from 32.9 to 47.6 over a single combined question, while raw-score SICR falls. Scoring all six dimensions cost about \$10 at list price.
\end{abstract}

\begin{figure}[t]
  \centering
  \includegraphics[width=\columnwidth]{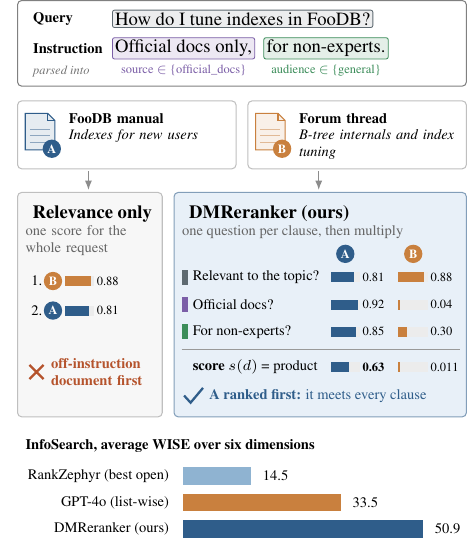}
  \caption{A relevance reranker puts the expert forum post first, because it is slightly more on topic. \sys{} asks one question per clause of the instruction and multiplies the answers, so the official manual for non-experts moves to the top. Each colour links a clause of the instruction to the question that checks it. Probabilities are illustrative. The bars show average WISE on InfoSearch, measured for \sys{} and as published for the others.}
  \label{fig:teaser}
\end{figure}

\section{Introduction}
\label{sec:intro}

A search query used to be a handful of keywords. Today it is often a sentence with conditions attached. A developer asks how to tune indexes in a database and adds that only the official documentation is useful. A student wants an explanation of photosynthesis that a twelve-year-old could follow. A reader wants news coverage in Spanish. In each case the user cares about two things at once, whether a document is about the right topic and whether it has the right properties. A document can be perfectly on topic and still be useless because it is a forum post, written for experts, or in the wrong language.

Rerankers are the natural place to handle such conditions, because they read the query and each candidate document together. Yet most rerankers were built to estimate topical relevance. Cross-encoders such as monoBERT and monoT5 are trained on relevance labels \citep{nogueira2019passage,nogueira2020monot5}, and large language model (LLM) rerankers prompted or distilled for ranking inherit the same goal \citep{sun2023rankgpt,pradeep2023rankvicuna,pradeep2023rankzephyr}. Recent benchmarks make the cost of this visible. FollowIR showed that retrieval models largely ignore instructions that change what counts as relevant \citep{weller2025followir}. InfoSearch went further and tested six document-level conditions, namely audience, keyword, format, language, length and source \citep{zhou2025infosearch}, and scored them with two metrics, WISE (Weighted Instruction Sensitivity Evaluation) and SICR (Strict Instruction Compliance Ratio). On InfoSearch, the strongest reranker reported, GPT-4o used list-wise, reaches an average WISE of 33.5, and the best open model reaches 14.5. Several rerankers score below zero, which means that adding the instruction moved the ranking in the wrong direction.

\begin{figure*}[t]
  \centering
  \includegraphics[width=\textwidth]{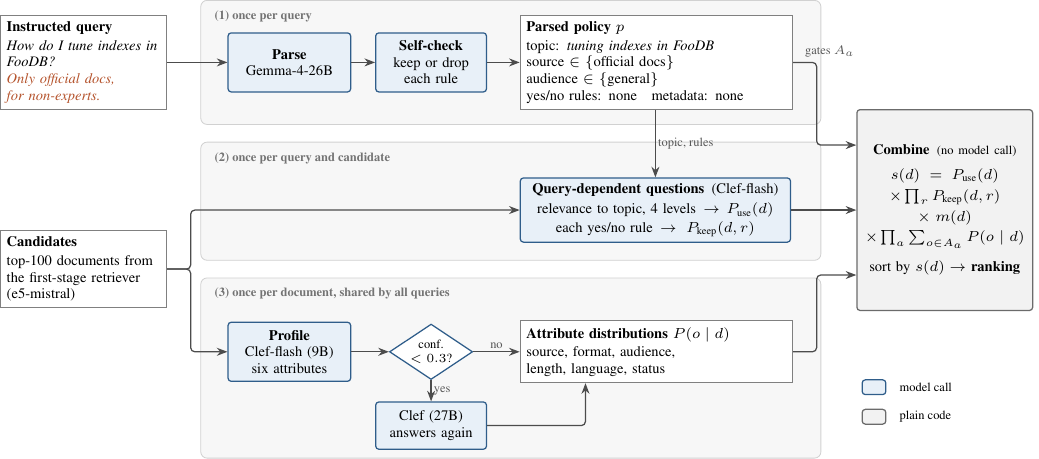}
  \caption{Architecture of \sys{}. (1)~An LLM parses the instructed query into a topic and typed clauses, and a second pass removes clauses it cannot support. (2)~For each query and candidate, a small judge model answers a four-level relevance question and one yes/no question per remaining rule. (3)~The same judge profiles every document once over six fixed attributes. When its confidence on an attribute that a rule needs is below 0.3, a larger judge answers again. Plain code multiplies the probabilities into the final score $s(d)$. Blue boxes call a model, the grey box does not.}
  \label{fig:arch}
\end{figure*}

We hypothesize that a large part of the problem is the form of the question that the model is asked. A list-wise reranker must weigh topic and conditions together and produce a single ordering, with no explicit place for a condition to veto a document. Our approach breaks the decision into pieces that we expect a model to answer more reliably. We first turn the instruction into a topic and a set of typed clauses. Then, for each candidate, we ask a judge model a series of narrow questions. Is this document relevant to the topic? Which audience is it written for? Is it official documentation? Each answer comes back as a probability. Finally we multiply the probabilities. A document that fails any clause receives a low score, no matter how relevant it is, and a document that satisfies every clause keeps its relevance score almost intact. Figure~\ref{fig:teaser} illustrates the idea on one query, and Figure~\ref{fig:arch} shows the full architecture.

This design has three practical benefits. It needs no training data, since each question is one that a general instruction-tuned judge already answers well. Document attributes do not depend on the query, so we profile each document once and reuse the profile for every query that touches it. Finally, every score can be traced back to the individual answers, so we can study what each component contributes by recombining saved answers, without calling a model again.

We describe \sys{} (Decision-Model Reranker), which parses an instruction into typed clauses and combines per-clause probabilities with a fixed product rule (Section~\ref{sec:method}). On InfoSearch it gives large gains in WISE and p-MRR over the published baselines on the same candidate pool, with no loss in plain relevance (Section~\ref{sec:results}). Section~\ref{sec:analysis} reports offline ablations, a comparison with one combined question to the same judge, the reason for its lower SICR, and the cost of the run.

\section{Related Work}
\label{sec:related}

\paragraph{Instruction following in retrieval.}
Early work on instructions in retrieval used them to describe the task, such as ``retrieve a passage that answers this question'' \citep{asai2023tart,su2023instructor}. Later benchmarks asked whether models follow instructions that change the meaning of relevance for a single query. FollowIR measures this with p-MRR, which compares rankings before and after the instruction is changed \citep{weller2025followir}. INSTRUCTIR attaches user-specific instructions to queries \citep{oh2024instructir}. Promptriever trains a bi-encoder retriever to respond to prompts \citep{weller2025promptriever}. InfoSearch, which we use, isolates six document-level conditions and evaluates each in an original, an instructed and a reversely instructed mode \citep{zhou2025infosearch}. Its two metrics, WISE and SICR, reward a model only when the ranking moves in the direction the instruction asks for.

\paragraph{LLM rerankers.}
LLMs can rerank point-wise, by scoring one document at a time, pairwise, or list-wise, by ordering a window of documents \citep{sun2023rankgpt,qin2024prp}. Setwise prompting reduces the calls of pairwise ranking \citep{zhuang2024setwise}, and Rank1 trains a reranker to reason before it answers \citep{weller2025rank1}. RankVicuna and RankZephyr distil list-wise ranking into open 7B models \citep{pradeep2023rankvicuna,pradeep2023rankzephyr}. Point-wise rerankers that read out label probabilities, rather than a single yes or no, give finer and better calibrated scores \citep{zhuang2024beyond}. Our relevance question follows this idea with a four-level scale. The difference is that we do not ask one model to fold the conditions into its relevance judgement. Each condition receives its own question and its own probability.

\paragraph{Decomposing the decision.}
Splitting a complex task into simpler steps is a common way to make LLM behaviour more reliable, both for reasoning \citep{zhou2023leasttomost,khot2023decomposed} and for evaluation, where a long text is checked one atomic fact at a time \citep{min2023factscore}. In retrieval, the closest analogue is filtering by metadata, which works when documents carry clean fields such as language or date. Web documents rarely do. Our document profile plays the role of these missing fields, filled in by a model and kept as probability distributions rather than hard labels, so that an uncertain attribute lowers a score without removing the document outright. Multiplying beliefs about separate conditions is old, as in the \texttt{\#and} operator of inference networks \citep{turtle1991inference}, soft conjunction in extended Boolean retrieval \citep{salton1983extended} and products of experts \citep{hinton2002poe}. What is new here is how the factors are obtained.

\section{Method}
\label{sec:method}

\sys{} takes an instructed query $q$ and a list of candidate documents and returns a score $s(d)$ for each candidate. It works in three steps. Only the first two call a model.

\subsection{Parsing the instruction}
\label{sec:parse}

A parser model (Gemma-4-26B-A4B, a mixture-of-experts model with about 4B active parameters \citep{google2026gemma4}) reads the query and returns a structured policy $p$ in JSON. The policy has four parts. The \emph{topic} is one sentence that states the information need with every condition removed. \emph{Attribute rules} restrict a document attribute to a set of allowed values, for example audience $\in\{$general$\}$. \emph{Other rules} cover conditions that do not fit an attribute. Each is a yes/no question about the document, together with the answer that keeps it, for example ``Does the document mention FooDB~3?'' with keep-if \emph{yes}. \emph{Metadata rules} cover dates and versions, such as ``published after 2023'' or ``most recent first''.

A second call to the same model checks the parse. It sees the query and the parsed rules and votes to keep or drop each rule, and it may rewrite the topic. It cannot add rules or change their content, so it can only make the policy more conservative. This step exists because a parser that invents a rule does more damage than one that misses a rule. An invented gate can push the only relevant document out of the top ten, while a missed rule leaves the ranking no worse than a plain relevance reranker.

\subsection{Asking the judge}
\label{sec:judge}

The judge is Clef-flash, a 9B model served on Cloudflare Workers AI that answers typed questions about a text and returns a probability distribution over the possible answers \citep{cloudflare2026clef}. It supports three question types that we use, a yes/no question, a choice among named options, and a score on an ordered scale.

\paragraph{Document profile.}
For each candidate document $d$ we ask six choice questions in one request, one per attribute: source, format, audience, length, language and status. Each attribute has a fixed set of options. The answer is a distribution $P(o \mid d)$ over the options of that attribute, together with a confidence value. These questions do not mention the query, so the profile of a document is computed once and shared by every query whose candidate list contains it. We wrote the attribute list with the six InfoSearch conditions in mind, and status was added for versioned documentation, so the profile fits this benchmark more closely than a list written for a new domain would. Its language options, for instance, are only English, Chinese and other. If a rule depends on an attribute whose confidence is below $\tau = 0.3$, the larger Clef model (27B) answers the profile again for that document, and its distribution replaces only the low-confidence attributes.

\paragraph{Relevance.}
We ask the judge how well $d$ addresses the parsed topic, on four levels from ``does not address it'' to ``fully answers it''. We use the probability that the level is at least 2,
\begin{equation}
  P_{\text{use}}(d) = P(\ell \ge 2 \mid d, \text{topic}).
\end{equation}
Asking about the parsed topic rather than the raw query matters. The raw query still contains the instruction, and a judge that sees ``for non-experts'' in the relevance question may fold the condition into its relevance estimate, which would count the condition twice.

\paragraph{Other rules.}
Each other rule $r$ becomes one yes/no question. If $P(\text{yes})$ is the judge's answer, the keep probability is $P_{\text{keep}}(d,r) = P(\text{yes})$ when the rule keeps on yes, and $1 - P(\text{yes})$ when it keeps on no.

\paragraph{Metadata.}
Date and version rules become yes/no questions in the same way, and ``most recent'' becomes a pairwise recency factor between candidates. We write the product of these factors as $m(d)$. InfoSearch does not grade documents by date, so $m(d) = 1$ for most queries. A few instructions still ask for ``the latest'' or ``current'' information, which the parser turns into a recency rule in 46 of the 4,791 query-mode rows (43 on Source). We kept these rules and measure their effect in Section~\ref{sec:ablation}.

\subsection{Combining the answers}
\label{sec:combine}

For an attribute rule $a$ with allowed set $A_a$, the probability that $d$ satisfies it is the mass the profile puts on the allowed options, $P_{\text{allowed}}(d,a) = \sum_{o \in A_a} P(o \mid d)$. The final score is
\begin{equation}
  \begin{split}
  s(d) \;=\; & P_{\text{use}}(d) \cdot \prod_{r} P_{\text{keep}}(d,r) \\
             & \cdot\, m(d) \cdot \prod_{a} P_{\text{allowed}}(d,a).
  \end{split}
  \label{eq:score}
\end{equation}
Equation~\ref{eq:score} treats the clauses as independent conditions that must all hold, so the score reads as the probability that the document is useful and satisfies every clause. There are no weights, thresholds or learned parameters. A document that clearly fails one clause gets a factor near zero and drops out of the top of the ranking. A document that clearly passes every clause keeps a score close to its relevance probability. Figure~\ref{fig:teaser} shows an example.

Every question, answer distribution and factor is written to a trace file. Since Equation~\ref{eq:score} is plain arithmetic over these saved values, we can drop or replace any factor afterwards and recompute the rankings without a single new model call. We use this for the ablations in Section~\ref{sec:ablation}.

\section{Experimental Setup}
\label{sec:setup}

\paragraph{Benchmark.}
InfoSearch has 600 core queries, 100 per dimension, and 6,392 documents \citep{zhou2025infosearch}. Each core query comes with instructed versions that add a condition, and reversely instructed versions that ask for the opposite. The paper reports 1,598 instructed queries. The public release contains 1,597, because one Keyword query is missing, and we evaluate on the release as is. We report all six dimensions, Audience (210 instructed queries), Keyword (287), Format (300), Language (200), Length (300) and Source (300).

\paragraph{Candidates.}
Following the InfoSearch reranking protocol, every system reranks the published top-100 candidates from e5-mistral \citep{wang2024e5mistral}, in all three modes. Because the candidates and relevance labels are identical, the reranker rows in our tables are same-pool comparisons. BM25 and e5-mistral rows are full-corpus retrieval results and are shown for reference only.

\paragraph{Metrics.}
We use the official InfoSearch evaluation code for WISE, SICR and p-MRR, and \texttt{pytrec\_eval} for nDCG@10 \citep{jarvelin2002ndcg}. WISE rewards a system when documents that satisfy the condition move up under the instruction and move down under the reverse instruction, with more weight on changes near the top of the ranking. SICR is stricter. It counts a query as followed only when the condition-satisfying document gains in the instructed mode and loses in the reversed mode. p-MRR, taken from FollowIR, measures the change in reciprocal rank between modes \citep{weller2025followir}. For each dimension we compute 95\% confidence intervals for our scores with a bootstrap over instructed queries (2,000 resamples) \citep{efron1993bootstrap,smucker2007significance}.

\paragraph{Baselines.}
We did not rerun any baseline. All baseline numbers come from Tables~1 and~2 of the InfoSearch paper, which uses the same candidates, labels and evaluation code \citep{zhou2025infosearch}. They cover point-wise rerankers (Mistral-7B-Instruct \citep{jiang2023mistral}, Llama-3.1-8B \citep{dubey2024llama3} and FollowIR-7B \citep{weller2025followir}) and list-wise rerankers (Mistral-7B-Instruct, Zephyr-7B-$\beta$ \citep{tunstall2023zephyr}, RankVicuna \citep{pradeep2023rankvicuna}, RankZephyr \citep{pradeep2023rankzephyr} and GPT-4o \citep{openai2024gpt4o}). InfoSearch reports p-MRR and nDCG@10 only as six-dimension averages, so we compare those as averages.

\paragraph{Implementation.}
The parser (\texttt{@cf/google/gemma-4-26b-a4b-it}) and both judges (\texttt{@cf/cloudflare/clef-flash} and \texttt{@cf/cloudflare/clef}) are called through Cloudflare Workers AI. All prompts, question wordings, the confidence threshold and the combination rule were fixed before the test runs (Appendix~\ref{app:prompts} gives their full text), and we record a hash of this configuration with every result. During development we tried the pipeline on 60 instructed queries each from Format and Source, which are also in the test set, and lowered $\tau$ from 0.5 to 0.3 to save cost. Without these 120 queries, WISE is 43.8 on Format and 67.3 on Source, against 45.3 and 65.2 on all queries. Responses are cached by request, so a rerun of the full evaluation makes no new calls and returns identical numbers.

\section{Results}
\label{sec:results}

\begin{table*}[t]
  \centering
  \small
  \setlength{\tabcolsep}{4.2pt}
  \begin{tabular}{l rrrrrr r}
    \toprule
    & Aud. & Key. & Form. & Lang. & Len. & Src. & Avg. \\
    \midrule
    \multicolumn{8}{l}{\emph{WISE}} \\
    BM25$^\dagger$                 & $-$3.0 & $-$42.1 & $-$2.8 & $-$7.2 & $-$7.5 & 2.0 & $-$12.0 \\
    e5-mistral$^\dagger$           & $-$7.3 & $-$44.5 & $-$19.9 & 0.1 & $-$13.4 & $-$13.0 & $-$16.3 \\
    Mistral-7B (point-wise)        & $-$8.9 & 34.5 & $-$9.3 & 21.4 & $-$12.6 & $-$0.3 & 4.1 \\
    Llama-3.1-8B (point-wise)      & $-$6.2 & 38.7 & $-$9.5 & 29.0 & $-$5.9 & 40.2 & 14.4 \\
    FollowIR-7B (point-wise)       & $-$2.3 & 47.7 & $-$2.3 & 20.9 & $-$2.6 & 18.8 & 13.4 \\
    Mistral-7B (list-wise)         & $-$6.3 & 46.0 & $-$6.6 & 7.6 & $-$1.9 & 10.0 & 8.1 \\
    Zephyr-7B-$\beta$ (list-wise)  & $-$2.7 & 14.1 & $-$13.9 & $-$6.9 & $-$5.7 & $-$3.9 & $-$3.2 \\
    RankVicuna (list-wise)         & $-$2.5 & $-$8.5 & $-$9.8 & $-$11.8 & $-$4.3 & $-$2.2 & $-$6.5 \\
    RankZephyr (list-wise)         & 7.4 & 53.9 & 7.8 & 10.6 & 7.8 & $-$0.3 & 14.5 \\
    GPT-4o (list-wise)             & 7.4 & \textbf{63.0} & 21.9 & \textbf{53.1} & 10.2 & 45.2 & 33.5 \\
    \rowcolor{cModelFill}\sys{} (ours)                  & \textbf{52.1} & 58.3 & \textbf{45.3} & 46.7 & \textbf{37.9} & \textbf{65.2} & \textbf{50.9} \\
    \midrule
    \multicolumn{8}{l}{\emph{SICR}} \\
    BM25$^\dagger$                 & 0.0 & 0.0 & 0.0 & 0.0 & 0.0 & 0.0 & 0.0 \\
    e5-mistral$^\dagger$           & 2.9 & 0.0 & 0.0 & 10.5 & 0.0 & 3.3 & 2.8 \\
    Mistral-7B (point-wise)        & 1.4 & 28.6 & 2.7 & 6.5 & 4.7 & 4.7 & 8.1 \\
    Llama-3.1-8B (point-wise)      & 6.2 & 38.3 & 10.0 & 22.0 & 2.7 & 36.7 & 19.3 \\
    FollowIR-7B (point-wise)       & 3.3 & 27.2 & 7.0 & 19.5 & 1.7 & 16.3 & 12.5 \\
    Mistral-7B (list-wise)         & 10.5 & 59.2 & 9.7 & 23.0 & 8.0 & 21.7 & 22.0 \\
    Zephyr-7B-$\beta$ (list-wise)  & 1.0 & 27.5 & 8.0 & 10.5 & 2.0 & 3.0 & 8.7 \\
    RankVicuna (list-wise)         & 5.2 & 10.5 & 3.3 & 4.5 & 2.3 & 8.0 & 5.6 \\
    RankZephyr (list-wise)         & 4.3 & 42.5 & 1.0 & 5.5 & 4.3 & 5.3 & 10.5 \\
    GPT-4o (list-wise)             & 15.2 & \textbf{60.6} & 11.3 & \textbf{55.5} & \textbf{10.3} & \textbf{39.7} & \textbf{32.1} \\
    \rowcolor{cModelFill}\sys{} (ours)                  & \textbf{19.5} & 43.6 & \textbf{14.0} & 20.5 & 7.7 & 10.0 & 19.2 \\
    \bottomrule
  \end{tabular}
  \caption{WISE and SICR per dimension on InfoSearch. All rerankers reorder the same e5-mistral top-100 candidates. Baseline values are from \citet{zhou2025infosearch}. ``Avg.'' is the six-dimension average. Bold marks the best same-pool reranker in each column. $^\dagger$Full-corpus retrieval, not a same-pool comparison.}
  \label{tab:main}
\end{table*}

\paragraph{Instruction following.}
Table~\ref{tab:main} gives the main comparison. Averaged over the six dimensions, \sys{} reaches a WISE of 50.9. GPT-4o reaches 33.5 and RankZephyr, the strongest open reranker by WISE, reaches 14.5. The 95\% bootstrap intervals for our WISE are [44.2, 59.6] on Audience, [52.5, 63.9] on Keyword, [38.1, 51.7] on Format, [38.0, 55.2] on Language, [31.6, 44.0] on Length and [59.3, 71.1] on Source. The baselines are published only as averages, so we cannot test the differences, and the comparisons below are between point estimates.

\begin{figure*}[t]
  \centering
\begin{tikzpicture}
  \pgfplotsset{
    resbars/.style={
      ybar, bar width=4.6pt, width=\textwidth, height=44mm,
      symbolic x coords={Audience,Keyword,Format,Language,Length,Source},
      xtick=data, enlarge x limits=0.09,
      ymin=-10, ymax=70, ytick={-10,0,20,40,60},
      ymajorgrids, grid style={black!12},
      axis line style={black!50}, tick style={black!50},
      tick label style={font=\rmfamily\scriptsize},
      label style={font=\rmfamily\footnotesize},
      title style={font=\rmfamily\footnotesize\bfseries, yshift=-1mm},
      extra y ticks={0}, extra y tick style={grid style={black!55}},
      legend style={font=\rmfamily\scriptsize, draw=none, legend columns=4,
                    /tikz/every even column/.append style={column sep=3mm}},
    },
  }
  \begin{axis}[resbars, title={WISE}, ylabel={score}, name=top,
               xticklabels={}, legend to name=resleg]
    \addplot[fill=cOurs, draw=none]   coordinates {(Audience,52.1) (Keyword,58.3) (Format,45.3) (Language,46.7) (Length,37.9) (Source,65.2)};
    \addplot[fill=cGPT, draw=none]    coordinates {(Audience,7.4) (Keyword,63.0) (Format,21.9) (Language,53.1) (Length,10.2) (Source,45.2)};
    \addplot[fill=cZephyr, draw=none] coordinates {(Audience,7.4) (Keyword,53.9) (Format,7.8) (Language,10.6) (Length,7.8) (Source,-0.3)};
    \addplot[fill=cFollow, draw=none] coordinates {(Audience,-2.3) (Keyword,47.7) (Format,-2.3) (Language,20.9) (Length,-2.6) (Source,18.8)};
    \legend{Ours, GPT-4o (list-wise), RankZephyr-7B, FollowIR-7B}
  \end{axis}
  \begin{axis}[resbars, title={SICR}, ylabel={score}, at={(top.south)}, anchor=north, yshift=-6mm]
    \addplot[fill=cOurs, draw=none]   coordinates {(Audience,19.5) (Keyword,43.6) (Format,14.0) (Language,20.5) (Length,7.7) (Source,10.0)};
    \addplot[fill=cGPT, draw=none]    coordinates {(Audience,15.2) (Keyword,60.6) (Format,11.3) (Language,55.5) (Length,10.3) (Source,39.7)};
    \addplot[fill=cZephyr, draw=none] coordinates {(Audience,4.3) (Keyword,42.5) (Format,1.0) (Language,5.5) (Length,4.3) (Source,5.3)};
    \addplot[fill=cFollow, draw=none] coordinates {(Audience,3.3) (Keyword,27.2) (Format,7.0) (Language,19.5) (Length,1.7) (Source,16.3)};
  \end{axis}
  \node[anchor=south] at ($(top.north)+(0,5mm)$) {\pgfplotslegendfromname{resleg}};
\end{tikzpicture}
  \caption{Per-dimension WISE (top) and SICR (bottom) on InfoSearch for \sys{} and three cited rerankers. \sys{} leads by a wide margin on Audience, Format, Length and Source WISE. GPT-4o leads on Keyword and Language WISE and on SICR overall. Baseline values are from \citet{zhou2025infosearch}.}
  \label{fig:results}
\end{figure*}
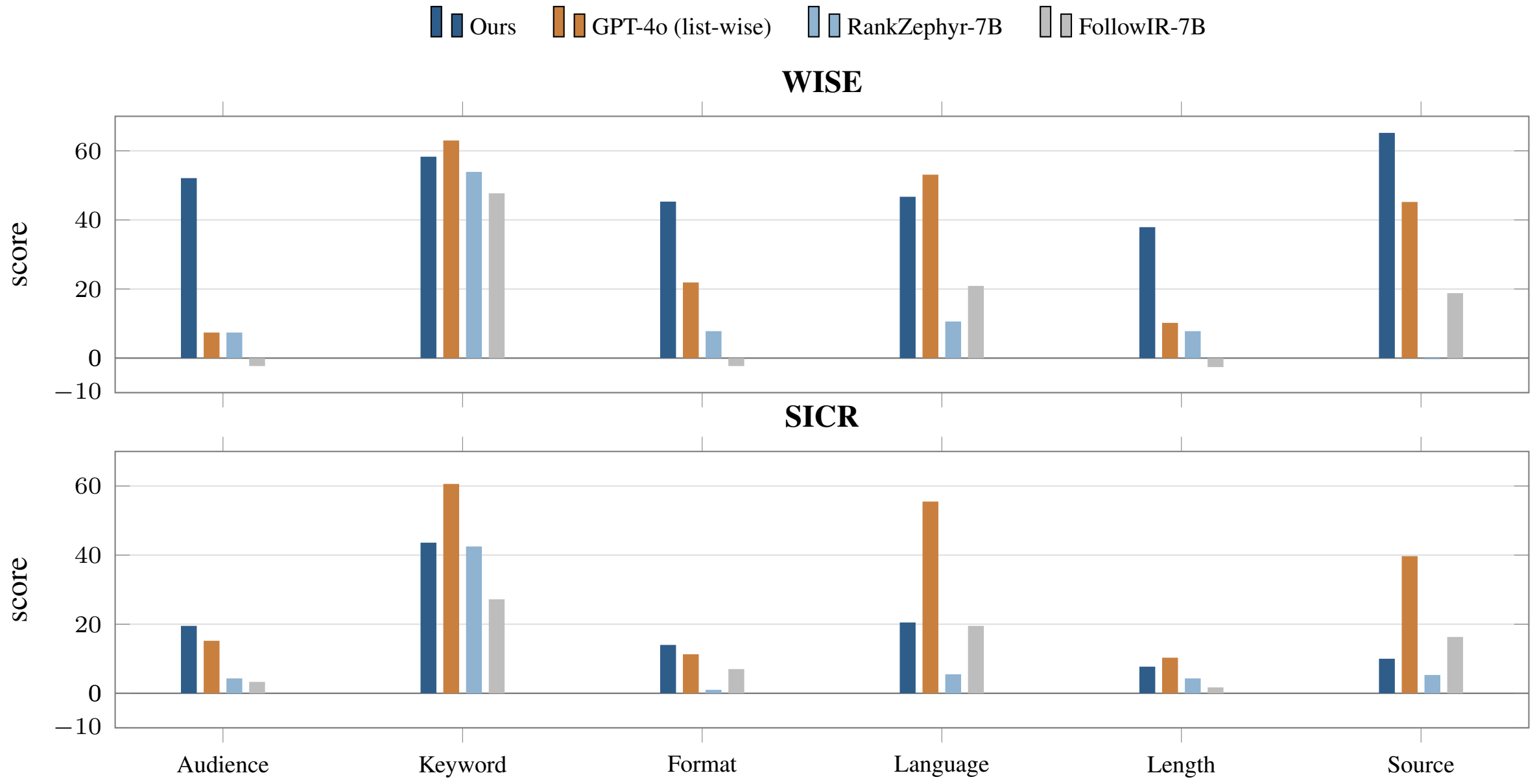

The average hides a clear pattern across dimensions, which Figure~\ref{fig:results} makes easy to see. The largest gains are on Audience (52.1 against 7.4 for GPT-4o), Format (45.3 against 21.9), Length (37.9 against 10.2) and Source (65.2 against 45.2). These are conditions on what kind of document something is, which most rerankers barely react to. Our document profile asks about them directly. On Keyword and Language, GPT-4o is ahead (63.0 against 58.3, and 53.1 against 46.7). Both are conditions that a strong LLM can check by reading the text, and GPT-4o has seen the whole list of candidates at once.

\paragraph{SICR.}
On SICR the picture is different. \sys{} is ahead of GPT-4o only on Audience and Format, and its average of 19.2 is below GPT-4o (32.1) and the list-wise Mistral-7B (22.0). Source, our best dimension by WISE, is one of our worst by SICR (10.0). Section~\ref{sec:sicr} explains why.

\begin{table}[t]
  \centering
  \small
  \setlength{\tabcolsep}{5pt}
  \begin{tabular}{l r rrr}
    \toprule
    & & \multicolumn{3}{c}{nDCG@10} \\
    \cmidrule(lr){3-5}
    & p-MRR & Ori. & Ins. & Rev. \\
    \midrule
    BM25$^\dagger$       & 7.0  & 47.5 & 39.1 & 38.5 \\
    e5-mistral$^\dagger$ & 4.0  & 78.3 & 64.3 & 66.0 \\
    FollowIR-7B (pw)     & 4.1  & 72.4 & 66.3 & 65.5 \\
    Llama-3.1-8B (pw)    & 11.5 & 74.8 & 66.8 & 65.4 \\
    RankZephyr (lw)      & 0.7  & 75.0 & 63.5 & 64.7 \\
    GPT-4o (lw)          & 15.0 & \textbf{83.8} & 74.2 & \textbf{74.2} \\
    \rowcolor{cModelFill}\sys{} (ours)        & \textbf{52.7} & 82.9 & \textbf{78.3} & 73.8 \\
    \bottomrule
  \end{tabular}
  \caption{p-MRR and nDCG@10 in the original, instructed and reversed modes. All values are six-dimension averages, as reported by InfoSearch. pw: point-wise, lw: list-wise. $^\dagger$Full-corpus retrieval.}
  \label{tab:pmrr}
\end{table}

\paragraph{Relevance is preserved.}
Table~\ref{tab:pmrr} shows that following instructions does not cost plain ranking quality. In the original mode, where there is no instruction, our nDCG@10 averages 82.9, close to GPT-4o (83.8) and above every open reranker in the InfoSearch table. In the instructed mode our nDCG@10 of 78.3 is above GPT-4o (74.2). Our p-MRR of 52.7 is far above the published values, which top out at 15.0 for GPT-4o.

\section{Analysis}
\label{sec:analysis}

\subsection{Which component does the work}
\label{sec:ablation}

Because Equation~\ref{eq:score} is computed from saved answers, we can remove a factor and recompute the rankings without new model calls. Table~\ref{tab:ablation} shows four variants, and Figure~\ref{fig:ablation} in Appendix~\ref{app:extra} breaks WISE down by dimension. With relevance only, WISE falls to about zero on every dimension ($-$4.0 on average), which confirms that the gains come from the clause factors and not from a better relevance model. Since the relevance question already uses the parsed topic, this variant is a strong relevance reranker that simply ignores the conditions.

The two kinds of clause split the work in an interpretable way. Removing the attribute gates collapses Audience ($-$1.2), Language ($-$9.1) and Source ($-$0.9), halves Format (20.6) and Length (17.6), but leaves Keyword unchanged at 58.3. Removing the yes/no rules does the opposite. Keyword drops to 4.1, while Audience and Language barely move and Source loses 4.7 points. On Keyword the parser keeps a required term in the topic, where the relevance question checks it, and turns an excluded term into a yes/no rule (286 of 287 reversed queries). Most other conditions go to attributes. Format and Length use both, which is why they lose ground under either removal.

We also checked the parse and the profile directly. The parser added no rule to any of the 1,597 original-mode queries. Against each document that the instruction makes irrelevant, the clause factors rank the document that meets it higher in 83.2\% of pairs in the instructed mode and 90.0\% in the reversed mode. Instructed Keyword is lowest (53.3\%), since its term stays in the topic, and the other dimensions range from 85.1\% (Length) to 99.0\% (Language).

The metadata factor is the one component that does not help. Removing it leaves five dimensions within 0.5 WISE points, but on Source it raises WISE from 65.2 to 67.6 and p-MRR from 74.8 to 81.7. The recency rules on Source come from instructions that ask for the latest information, so the parser read them correctly. InfoSearch, however, labels Source documents by the kind of source and not by date, and reordering candidates by date adds noise to that judgement. The factor was also the most expensive part of the run (Section~\ref{sec:cost}). We report the frozen system as it was run and do not remove the factor after the fact.

\begin{table*}[t]
  \centering
  \small
  \begin{tabular}{l rrr rr}
    \toprule
    Variant & WISE & SICR & p-MRR & nDCG@10 Ins. & nDCG@10 Rev. \\
    \midrule
    \rowcolor{cModelFill}V0\enspace Full system                 & 50.9 & 19.2 & 52.7 & 78.3 & 73.8 \\
    V1\enspace Relevance only               & $-$4.0 & 4.2 & 1.6 & 64.8 & 71.0 \\
    V2\enspace No metadata factor           & \textbf{51.3} & 19.4 & 53.9 & \textbf{78.5} & 73.9 \\
    V3\enspace No attribute gates           & 14.2 & 15.0 & 2.2 & 65.0 & 70.7 \\
    V4\enspace No yes/no rules              & 34.9 & 12.3 & 52.3 & 78.0 & \textbf{74.6} \\
    \midrule
    V5\enspace Mean instead of product (post hoc) & 47.4 & \textbf{33.0} & \textbf{57.2} & 77.2 & 73.5 \\
    \bottomrule
  \end{tabular}
  \caption{Offline ablations, averaged over the six dimensions. Each variant recombines the saved answers of the full system, so no new model calls were made. V5 was tried after seeing test results and is not part of the system.}
  \label{tab:ablation}
\end{table*}

\subsection{Clause questions against one combined question}
\label{sec:single}

The ablations show that the clause factors carry the gains, but they do not show that splitting the instruction is better than asking about it directly. Relevance only ignores the instruction, and the published baselines use other models. We therefore built a single-question baseline on the same judge. For each query and candidate, Clef-flash answers one yes/no question, whether the document is useful for the query including its instruction, and the probability of yes is the score. There is no parse, no profile and no product, and the question is yes/no rather than our four-level scale.

To keep the cost small we ran this baseline on a subset of 50 instructed queries per dimension, 300 in total, drawn with a fixed random seed. The subset was fixed and recorded before any baseline call was made. Both systems are scored on exactly these queries in all three modes, and we compare them with a paired bootstrap over queries (2,000 resamples). The baseline needed 85,170 judge calls and cost \$1.22 at list price.

\begin{table}[t]
  \centering
  \small
  \setlength{\tabcolsep}{4pt}
  \begin{tabular}{l rr r c}
    \toprule
    Metric & Single & Ours & Diff. & 95\% CI \\
    \midrule
    WISE          & 32.9 & \textbf{47.6} & $+$14.7 & [7.3, 22.2] \\
    SICR          & \textbf{37.0} & 20.3 & $-$16.7 & [$-$22.7, $-$10.3] \\
    p-MRR         & 20.0 & \textbf{53.0} & $+$33.0 & [28.4, 38.1] \\
    nDCG@10 Ori.  & 77.9 & \textbf{81.6} & $+$3.7  & [1.4, 6.0] \\
    nDCG@10 Ins.  & 72.5 & \textbf{76.7} & $+$4.1  & [0.8, 7.5] \\
    nDCG@10 Rev.  & 66.8 & \textbf{74.0} & $+$7.2  & [3.9, 10.4] \\
    \bottomrule
  \end{tabular}
  \caption{\sys{} against a single combined question to the same judge (Clef-flash), on a fixed random subset of 300 instructed queries, 50 per dimension. ``Single'' is the single question and ``Ours'' is \sys{}. Differences are \sys{} minus the single question, with paired bootstrap 95\% intervals.}
  \label{tab:single}
\end{table}

Table~\ref{tab:single} gives the result. Splitting the instruction raises WISE from 32.9 to 47.6 and p-MRR from 20.0 to 53.0, and both intervals exclude zero. \sys{} also has better nDCG@10 in all three modes, but in the original mode, which has no clauses, the gain comes from the four-level question and the parsed topic, not from splitting. On SICR the result reverses, and the single question is clearly ahead (37.0 against 20.3). Per dimension, the WISE gain is large on Audience ($+$46.9) and Format ($+$25.6) and within noise elsewhere, with the single question ahead on Source by 10.5 points. With 50 queries per dimension these intervals are wide, so we treat the macro difference as the main result.

\subsection{Why SICR lags}
\label{sec:sicr}

SICR asks whether the score of a document that satisfies the condition rises when the instruction is added. In our model every clause factor lies between 0 and 1, so adding a clause can never raise a score by itself. A matching document only gains if its relevance probability goes up, and that rarely happens, because the parser strips the condition from the topic and the topic is therefore nearly the same in both modes. Attribute gates help the matching document by pushing the other documents down, which is exactly what WISE and p-MRR measure, but its own score usually stays the same or falls a little. On Language, 159 of the 200 instructed queries have an SICR of zero, and in most of them the instructed-mode score of the matching document is slightly lower than its original-mode score.

Two further results support this reading. The single question asks about the instruction directly, so a matching document can gain score, and its SICR is much higher than ours (Section~\ref{sec:single}). The post-hoc variant V5 (Table~\ref{tab:ablation}) replaces the product with a mean, which lets a passed clause add to the score. SICR rises from 19.2 to 33.0 while WISE falls from 50.9 to 47.4. V5 was chosen after seeing test results, so we report it only as evidence about the metric.

The official code compares the raw scores of the matching document across the three runs. InfoSearch does not say how scores were assigned to the output of list-wise rerankers. If they come from ranks, SICR reduces to its rank conditions, which a gate that pushes other documents down does meet. Scoring our runs by rank raises our SICR from 19.2 to 67.0, and on the 300-query subset it reverses the comparison with the single question (67.0 against 52.0). SICR values are thus comparable only between systems whose scores are formed the same way.

\subsection{Cost}
\label{sec:cost}

Scoring all six dimensions took 695,393 Clef-flash calls and about 274 million input tokens, which comes to about \$10.4 at the Workers AI list price, or about two thirds of a cent per instructed query. Appendix~\ref{app:cost} breaks this down by dimension (Table~\ref{tab:cost}).

We did not run GPT-4o, but counting the tokens of the actual candidates under the sliding-window setup of InfoSearch and pricing them at the GPT-4o list price gives about \$486 for the same 4,791 rankings, roughly 47 times our cost (Appendix~\ref{app:cost}). The two prices come from different providers, so the ratio is only a rough guide.

Most of our cost comes from one component. A ``most recent'' rule compares every pair of the 100 candidates in both orders, 9,900 calls per query. The 43 Source rows with such a rule asked 425,700 pair questions, 415,800 of them unique because two rows share their topic and candidates. That is 60\% of all judge calls, and Section~\ref{sec:ablation} shows they did not improve the ranking.
\section{Conclusion}
\label{sec:conclusion}

We showed that a reranker can follow document-level instructions well without any training, by turning the instruction into small questions that a judge model can answer and multiplying the answers. The clause questions carry the gains, and on the same judge, splitting the instruction raises WISE from 32.9 to 47.6 over one combined question, although raw-score SICR falls because the product rule cannot raise a score. On InfoSearch, this simple design scores well above the published GPT-4o result on average WISE and on four of the six dimensions, keeps plain relevance intact, and costs under one cent per instructed query. We think the same pattern (parse once, profile once, ask narrow questions and combine in code) is worth trying wherever the conditions users state can be named in advance as document attributes.

Fine-tuning has improved instruction following in retrieval models \citep{weller2025followir,weller2025promptriever,zhou2025infosearch}, and training our judge on attribute and rule questions, for example with labels from a larger model, could sharpen its answers.

\section*{Limitations}
\label{sec:limits}

The single-question comparison covers a random subset of 300 instructed queries, not the full benchmark, so its per-dimension results are noisy. Apart from that baseline, our baseline numbers are cited from the InfoSearch paper and not reproduced. The candidate pools, labels and evaluation code are the same, but we cannot rule out small differences in evaluation details. The public release has one Keyword query fewer than the paper, so our Keyword average is over 287 queries rather than 288.

The system depends on hosted models. We pin model identifiers and cache every response, which makes our numbers reproducible from the cache, but a provider may update a hosted model in place. The judge has open weights, and a local rerun would remove this dependence.

The parser can add rules that the benchmark does not grade. On InfoSearch it added a recency rule to 46 query-mode rows. Each of these matched wording in the instruction, so none was a parse error, but InfoSearch does not grade recency, and on Source the factor lowered scores slightly while adding most of the cost. A parser that invents a rule could do more harm, since a wrong gate can remove a relevant document. The self-check step reduces this risk without removing it.

We evaluate on a single benchmark. InfoSearch conditions map cleanly onto document attributes, which suits our design, and our attribute list was written with them in mind. A new domain needs its own list, and we have not measured how much a list written without the benchmark in view would lose. Instructions that redefine relevance in subtler ways, as in FollowIR, may need more yes/no rules and fewer attribute gates. A first attempt to reproduce the FollowIR candidate pool did not match the published first-stage numbers, so we leave FollowIR to future work rather than report numbers on a pool we cannot verify. Finally, the method needs one judge call per candidate and question. From an empty cache a ranking needed 145 judge calls on average, or 58 without the Source recency pairs, where GPT-4o needs 9 list-wise windows. The calls are small and independent, so they can run in parallel and latency depends more on the provider's concurrency than on their number. We did not measure latency.

\bibliography{refs}

\clearpage
\appendix
\renewcommand{\dbltopfraction}{0.8}
\renewcommand{\dblfloatpagefraction}{0.95}

\section{Prompts and Questions}
\label{app:prompts}

This appendix gives the exact text of every prompt and question in the frozen configuration. Placeholders in braces are filled at run time.

\paragraph{Parser.}
Figure~\ref{fig:prompt-parse} shows the parse prompt and Figure~\ref{fig:prompt-check} the self-check prompt. Both go to Gemma-4-26B-A4B with the system message ``You convert search requests into structured filters. You answer with JSON only.'' The parse call allows 700 output tokens and the self-check 400. The first attempt uses temperature 0. If the reply is not valid JSON of the expected shape, the call is repeated at temperature 0.4, at most three times in total. If the self-check fails all three times, the parse is kept unchanged. The \texttt{\{taxonomy\}} placeholder lists the six attributes with their option names (Table~\ref{tab:taxonomy}), one line per attribute. An empty instruction is shown as ``(none)''.

\paragraph{Judge.}
Clef-flash does not take a free-form prompt. Each request carries the document text and a set of typed questions, and the model returns a probability for every answer. Table~\ref{tab:questions} lists the questions we ask. The profile questions are sent as one request per document, always in the same order. All yes/no rule questions for a document and query are also sent as one request.

\paragraph{Recency.}
For a ``most recent'' rule the judge sees two documents, A and B, each cut to its first 80 words, and answers the choice question in Table~\ref{tab:questions}. Each pair is asked in both orders, and the two answers are averaged to cancel position bias, which gives $P(e \succ d)$, the probability that $e$ is more recent than $d$. The factor for $d$ is $\prod_{e \neq d} \bigl(1 - P(e \succ d)\, P_{\text{use}}(e)\bigr)$, the probability that no other useful candidate is more recent.

\begin{table*}[t]
  \centering
  \small
  \begin{tabular}{l l p{0.62\textwidth}}
    \toprule
    Attribute & Option & Description given to the judge \\
    \midrule
    source   & official\_docs & Official documentation, manual or reference published by the project, vendor or organisation \\
             & blog           & A blog post (personal or company blog) \\
             & forum\_qa      & A forum thread or Q\&A site post, e.g. Stack Overflow, Reddit, Quora \\
             & news           & A news article from a news outlet or newspaper \\
             & academic       & A scientific paper, study or academic text \\
             & other          & Anything else, e.g. encyclopedia, product page, generic web page \\
    \midrule
    format   & prose          & Mainly explanatory prose \\
             & code           & Mainly a code snippet with little or no prose \\
             & qa\_post       & A question followed by one or more answers \\
             & list\_table    & Mainly a list, table or bullet points \\
    \midrule
    audience & expert         & Written for experts or specialists, using technical language \\
             & general        & Written for a general, non-expert audience \\
    \midrule
    length   & sentence       & A single sentence or very short snippet \\
             & paragraph      & One paragraph \\
             & article        & Several paragraphs or a full article \\
    \midrule
    language & english        & Written in English \\
             & chinese        & Written in Chinese \\
             & other          & Written in another language \\
    \midrule
    status   & current        & Describes current, recommended practice \\
             & deprecated     & Describes something marked deprecated, removed, superseded or outdated \\
             & unclear        & Gives no indication either way \\
    \bottomrule
  \end{tabular}
  \caption{The six profile attributes and their options. The parser sees the option names, and the judge sees the descriptions.}
  \label{tab:taxonomy}

  \vspace{5mm}
  \begin{tabular}{l l p{0.56\textwidth}}
    \toprule
    Step & Type & Question and answer options \\
    \midrule
    Profile (per attribute) & choice & ``Which best describes the document's \{attribute\}?'' Options: the descriptions in Table~\ref{tab:taxonomy}. \\
    Relevance & score & ``How completely does this document answer this need: "\{topic\}"?'' Levels 0--3: Does not address it; Mentions the subject but gives no usable information; Partially answers it; Fully answers it. \\
    Other rule & yes/no & The question written by the parser, e.g. ``Does the document discuss the flu?'' \\
    Date rule & yes/no & ``According to this document's information, was it last modified or published on or after \{date\}?'' (or ``before \{date\}?'') \\
    Version rule & yes/no & ``Is this document about version \{version\}?'' \\
    Recency pair & choice & ``Which of the two documents contains the more recent, up-to-date information about: "\{topic\}"?'' Options: Document A, Document B. \\
    \bottomrule
  \end{tabular}
  \caption{Questions sent to the Clef-flash judge. Dates and versions are passed exactly as the parser wrote them.}
  \label{tab:questions}

  \vspace{5mm}
  \begin{minipage}{0.6\textwidth}
    \centering
    \setlength{\tabcolsep}{4pt}
    \begin{tabular}{l rrrr}
      \toprule
      Dimension & Calls & Tok.\ (M) & USD & \textcent/query \\
      \midrule
      Audience & 48,743  & 24.6  & 0.93 & 0.44 \\
      Keyword  & 30,695  & 6.8   & 0.26 & 0.09 \\
      Format   & 45,025  & 17.6  & 0.67 & 0.22 \\
      Language & 43,877  & 14.9  & 0.56 & 0.28 \\
      Length   & 59,166  & 14.5  & 0.55 & 0.18 \\
      Source   & 467,887 & 196.1 & 7.45 & 2.48 \\
      \quad recency pairs & 415,800 & & & \\
      \midrule
      Total    & 695,393 & 274.4 & 10.42 & 0.65 \\
      \midrule
      \multicolumn{5}{l}{\emph{GPT-4o list-wise, estimated, not run by us}} \\
      Total    & 43,119  & 180.2 & 486 & 30.4 \\
      \bottomrule
    \end{tabular}
    \caption{Judge calls, input tokens and cost at list price. \textcent/query covers the three rankings of one instructed query. Length and Source tokens are reconstructed after an interrupted run (about 15\% accuracy). GPT-4o tokens include 4.7M output tokens.}
    \label{tab:cost}
  \end{minipage}
\end{table*}

\section{Per-Dimension Ablations}
\label{app:extra}

\begin{figure}[!h]
  \centering
\begin{tikzpicture}
  \begin{axis}[
      ybar, bar width=2.3pt, width=\columnwidth, height=28mm,
      symbolic x coords={Audience,Keyword,Format,Language,Length,Source},
      xtick=data, enlarge x limits=0.09,
      ymin=-15, ymax=70, ytick={0,20,40,60}, ylabel={WISE},
      ymajorgrids, grid style={black!12},
      extra y ticks={0}, extra y tick style={grid style={black!55}},
      axis line style={black!50}, tick style={black!50},
      tick label style={font=\rmfamily\scriptsize},
      label style={font=\rmfamily\footnotesize},
      legend style={font=\rmfamily\scriptsize, draw=none, at={(0.5,1.03)},
                    anchor=south, legend columns=2, cells={anchor=west},
                    /tikz/every even column/.append style={column sep=3mm}},
    ]
    \addplot[fill=cOurs, draw=none]   coordinates {(Audience,52.1) (Keyword,58.3) (Format,45.3) (Language,46.7) (Length,37.9) (Source,65.2)};
    \addplot[fill=cAblA, draw=none]   coordinates {(Audience,50.2) (Keyword,4.1) (Format,33.8) (Language,46.7) (Length,14.1) (Source,60.5)};
    \addplot[fill=cAblB, draw=none]   coordinates {(Audience,-1.2) (Keyword,58.3) (Format,20.6) (Language,-9.1) (Length,17.6) (Source,-0.9)};
    \addplot[fill=cAblC, draw=none]   coordinates {(Audience,-2.8) (Keyword,4.1) (Format,-3.7) (Language,-9.7) (Length,-3.0) (Source,-9.1)};
    \legend{Full system, No yes/no rules, No attribute gates, Relevance only}
  \end{axis}
\end{tikzpicture}
  \caption{WISE per dimension when one component is removed. Attribute gates carry Audience, Language and Source. Yes/no rules carry Keyword. Format and Length need both. Relevance alone does not follow instructions on any dimension.}
  \label{fig:ablation}
\end{figure}
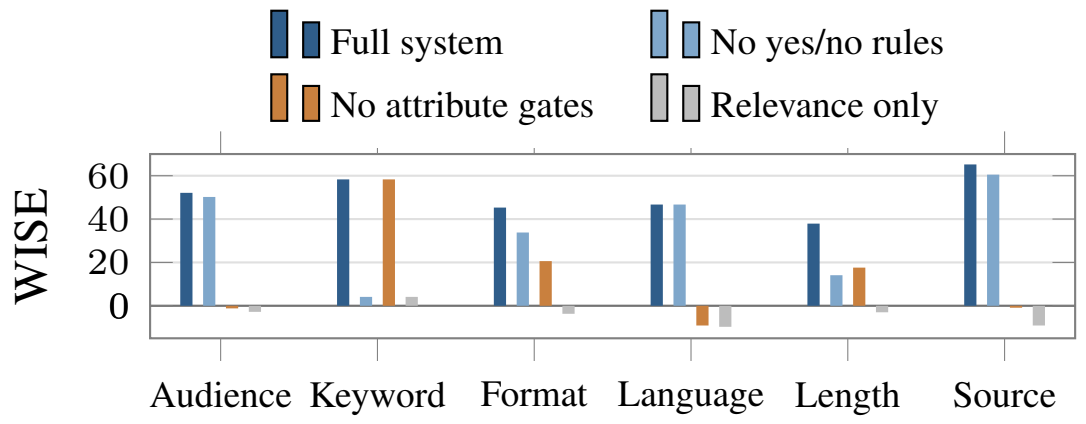

Figure~\ref{fig:ablation} shows the WISE of each ablation variant per dimension. The averages are in Table~\ref{tab:ablation}, and Section~\ref{sec:ablation} discusses the pattern.

\section{Cost Details}
\label{app:cost}

Table~\ref{tab:cost} lists the judge calls and their cost per dimension. Calls are the unique requests a dimension needs from an empty cache, priced at \$0.038 per million input tokens. The parser added 4,791 calls, one parse and one self-check per query-mode row. These and the calls to the larger Clef model are not priced in the table.

\begin{figure*}[!t]
\begin{Verbatim}[fontsize=\scriptsize, frame=single, breaklines, breakanywhere=false]
Search request: {query}
Instructions / requirements from the user: {instruction}

Documents have these attributes (attribute: allowed option names):
{taxonomy}
Documents may also carry metadata: a last-modified date and a version.

Convert the request into JSON:
- "topic": one sentence describing the information the user needs. Do NOT include any rule: no source, format, audience, length, language, status, date, recency or version.
- "metadata_rules": date, recency and version requirements ONLY, using these forms:
    {"type": "date_on_or_after", "date": "YYYY-MM-DD"}, {"type": "date_before", "date": "YYYY-MM-DD"},
    {"type": "most_recent"}  (user wants only the newest / most up-to-date information),
    {"type": "version_equals", "version": "<exact version string>"}
- "attribute_rules": requirements or exclusions about the attributes above that the user EXPLICITLY mentions, with the list of option names that are ALLOWED. Example: "not a blog" -> {"attribute": "source", "allowed": ["official_docs", "forum_qa", "news", "academic", "other"]}.
- "other_rules": any other requirement or exclusion, as a simple positive yes/no question plus "keep_if" ("yes" or "no"). Example: "ignore documents about the flu" -> {"question": "Does the document discuss the flu?", "keep_if": "no"}.
- Only encode requirements the user explicitly stated; never invent rules. Express each requirement exactly once (dates/versions/recency only in metadata_rules).
Use empty lists when there are no such rules.
Return only JSON: {"topic": "...", "metadata_rules": [...], "attribute_rules": [...], "other_rules": [...]}
\end{Verbatim}
\caption{Parse prompt for Gemma-4-26B-A4B.}
\label{fig:prompt-parse}

\vspace{4mm}
\begin{Verbatim}[fontsize=\scriptsize, frame=single, breaklines, breakanywhere=false]
Search request: {query}
Instructions / requirements from the user: {instruction}

A parser turned this into a topic and a numbered list of rules:
Topic: {topic}
Rules:
{rules}

Only the line "Instructions / requirements from the user" contains instructions.
For each rule answer true (keep) or false (remove):
- true if the instructions ask for it. The parser may have reworded it, mapped it to option names or written it as
  a yes/no question; that still counts as asked for. Examples: instructions "Only use the most recent information"
  -> {"type": "most_recent"} is true; instructions "Only consider official documents" -> {"attribute": "source",
  "allowed": ["official_docs"]} is true; instructions "Do not use forums or blogs" -> {"attribute": "source",
  "excluded": ["forum_qa", "blog"]} is true.
- false if the instructions do not ask for it: e.g. the instructions are (none), or a source rule when the
  instructions only name a version.
- false if it repeats a requirement already expressed by an earlier rule in the list.
When in doubt, answer true.
Also rewrite the topic so it describes only the information need, without any version, date, recency, source,
format, audience, length or language requirement.
Return only JSON: {"topic": "...", "keep": [true/false for rule 1, rule 2, ...]} with exactly {n} entries in "keep".
\end{Verbatim}
\caption{Self-check prompt. \texttt{\{rules\}} is a numbered list of the parsed rules in JSON. An attribute rule that allows most options is shown by the options it excludes, so ``not a blog'' reads as \texttt{"excluded": ["blog"]}. This changes only the display, not the rule.}
\label{fig:prompt-check}
\end{figure*}

\paragraph{GPT-4o estimate.}
We did not run GPT-4o. InfoSearch follows the sliding-window setup of RankZephyr \citep{pradeep2023rankzephyr,sun2023rankgpt}, which ranks the top 100 candidates in 9 windows of 20 with a stride of 10. For each of the 4,791 rankings, we counted the tokens of the actual candidate texts with the GPT-4o tokenizer (\texttt{o200k\_base}), added 150 prompt tokens per window for the template and the query, and assumed 110 output tokens per window for the ranked list of identifiers. Windows overlap, so each ranking reads 180 candidate slots, which we priced at the mean candidate length of that ranking. This gives 43,119 calls, 175.5 million input tokens and 4.7 million output tokens. At the GPT-4o list price of \$2.50 per million input and \$10 per million output tokens, the total is about \$486.

\paragraph{Where the cost goes.}
Apart from the recency pairs, cost follows document length more than query count. Since profiles are computed once per document, a new query adds only its relevance and rule questions. A single choice question over the whole list, or sorting on an extracted date, would remove nearly all of the extra cost on Source.

\end{document}